**Y. Yao, K. Mizuno, K. Ohnishi, Y. Ishikawa, M. Kitahara, T. Tomida, R. Murakami, V. Kochurikhin, L. Gushchina, K. Kamada, K. Kakimoto, and A. Yoshikawa**

**Synchrotron-radiation x-ray topography and reticulography of bulk β-$Ga_2O_3$ crystals grown by the cold crucible method**

# Synchrotron-radiation X-ray topography and reticulography of bulk β-$Ga_2O_3$ crystals grown by the cold crucible method

Yongzhao Yao,[1,2,a)] Koki Mizuno,[1] Kazuki Ohnishi,[1] Yukari Ishikawa,[2] Masanori Kitahara,[3,4] Taketoshi Tomida,[3] Rikito Murakami,[4] Vladimir Kochurikhin,[3] Liudmila Gushchina,[3] Kei Kamada,[3,5] Koichi Kakimoto,[4] Akira Yoshikawa[3,4,5]

[1]Mie University, 1577 Kurimamachiya-cho, Tsu, Mie 514-8507, Japan
[2]Japan Fine Ceramics Center, 2-4-1 Mutsuno, Atsuta, Nagoya 456-8587, Japan
[3]C&A Corporation, 1-16-23 Ichibanchho, Aoba-ku, Sendai, Miyagi 980-0811, Japan
[4]Institute for Materials Research, Tohoku University, 2-1-1, Katahira, Aoba-ku, Sendai, Miyagi 980-8577, Japan
[5]New Industry Creation Hatchery Center, Tohoku University, 6-6-10, Aramaki Aza Aoba, Aoba-ku, Sendai, Miyagi 980-8579, Japan

**Abstract:** The structural properties of a β-$Ga_2O_3$ single crystal grown by the oxide crystal growth from cold crucible (OCCC) method were investigated using synchrotron radiation X-ray topography and X-ray reticulography. The region grown beneath the seed exhibits high crystalline quality with a rocking curve full width at half maximum of about 26 arcsec. During diameter enlargement, a twist-type lattice misorientation develops between the central and laterally expanded regions, originating near the shoulder and propagating along boundaries parallel to the ⟨010⟩ growth direction. Dislocation analysis reveals that ⟨010⟩-oriented screw dislocations dominate the defect structure with densities of ~$10^5$ $cm^{-2}$, while higher densities (~$10^6$ $cm^{-2}$) appear in the wing region (i.e., the laterally expanded region formed during diameter enlargement). These results clarify defect formation in OCCC-grown β-$Ga_2O_3$ and provide insights for optimizing growth conditions.

[a)]Author to whom correspondence should be addressed. Electronic mail: yao@icsdf.mie-u.ac.jp.
ORCID:0000-0002-7746-4204

## I. INTRODUCTION

β-$Ga_2O_3$ has emerged as a highly promising ultra-wide-bandgap (UWBG) semiconductor for next-generation high-power and high-temperature electronic devices due to its large bandgap (~4.8–4.9 eV), high critical breakdown field (>8 MV/cm), and the unique availability of bulk single crystals grown from the melt. These advantages enable vertical device architectures with extremely high Baliga's figure of merit, placing β-$Ga_2O_3$ alongside SiC and GaN as a leading material for power electronics [1, 2, 3, 4]. However, the performance and long-term reliability of β-$Ga_2O_3$ devices depend critically on the structural quality of the bulk substrates. Extended defects—particularly dislocations, sub-grain boundaries, and pipe-shaped voids—strongly influence subsequent epitaxial layer quality, carrier transport, and breakdown behavior [5, 6, 7, 8, 9, 10]. As device performance becomes increasingly substrate-limited, understanding how melt growth techniques generate specific defect populations has become an essential research focus.

A variety of melt-based techniques have been explored for producing β-$Ga_2O_3$, each offering a different balance of growth stability, throughput, contamination control, and defect formation. Among crucible-based methods, the edge-defined film-fed growth (EFG) method provides a pathway toward industrial wafer production by enabling continuous growth and large cross-sectional areas [11]. However, EFG requires iridium (Ir) crucibles, a noble metal that is both expensive [12] and unstable under high oxygen partial pressures at elevated temperatures. As a result, significantly reducing the cost of EFG substrates is difficult, and Ir-related contamination, in addition to certain characteristic point defects arising from the relatively low oxygen partial pressure, is intrinsic to the EFG process [13, 14]. In addition, EFG is limited to crystal growth along the ⟨010⟩ direction. Common defect types include pipe-like nanovoids [5], slip-activated dislocation arrays, and domain boundaries, as revealed by X-ray topography [15, 16, 17, 18, 19] and other characterization techniques [20, 21, 22, 23, 24, 25]. Because the defect structure strongly depends on the growth method, detailed characterization of defects is essential for understanding and comparing different crystal growth techniques. Other melt growth methods for β-$Ga_2O_3$ reported in the literature include the Czochralski (CZ) method [26, 27, 28, 29, 30], the vertical Bridgman (VB) method [31, 32, 33, 34, 35, 36], and the casting method [37, 38, 39]. Similar to EFG, all of these methods require the use of noble-metal crucibles such as Ir, Pt, and Rh. Estimates have shown that more than half of the total β-$Ga_2O_3$ wafer cost is attributable to the extremely high cost of the crucible when Ir is used [12].

To reduce wafer cost, the oxide crystal growth from cold crucible (OCCC) method, which does not require crucibles made of noble metals, has been proposed for the bulk growth of β-$Ga_2O_3$ [40, 41]. The apparatus consisted of a high-frequency oscillation coil, a water-cooled copper (Cu) basket with internal cold-water circulation, and a seed pulling and rotation mechanism. OCCC utilizes both a conventional high-frequency (~5 MHz) vacuum tube generator and an original low-frequency (0.4–0.5 MHz) SiC MOSFET generator, which provides improved efficiency and control. In this system, the temperature gradient is controlled externally rather than by container heating, in contrast to conventional furnace-based methods where input power directly determines the temperature. This configuration enables monitoring of the melt state through changes in electrical resistance and allows stable heat generation via electromagnetic feedback. The process of melt stabilization was carefully controlled by adjusting heating parameters such as the oscillation frequency of the SiC MOSFET transistor generator [40]. In this process, the sintered high-purity raw material itself acted as the container [42]. Owing to the low temperature near the Cu basket walls, a thin unmelted layer formed and functioned as a self-supporting crucible for the melt. After melt stabilization, crystal growth proceeded in a manner similar to that of the conventional CZ method. Further details of OCCC growth are provided in Ref. 40. The growth rates were typically controlled between 3 and 15 mm/h. A high-crystalline-quality bulk β-$Ga_2O_3$ crystal grown by the OCCC method, with a diameter of 32 mm and a length of 50.8 mm, and exhibiting an X-ray diffraction (XRD) full width at half maximum (FWHM) of 23.6 arcsec, has been demonstrated [41]. Because there is no concern regarding reactions between crucible and oxygen, a wide range of oxygen partial pressures, up to a pure oxygen atmosphere, can be applied in the OCCC method. The use of a high oxygen partial pressure not only mitigates the decomposition of molten $Ga_2O_3$ and the formation of elemental Ga, but is also expected to reduce both extended defects and point defects arising from insufficient oxygen supply.

Although the OCCC method has demonstrated great potential as a bulk growth method for β-$Ga_2O_3$, the defect landscape of OCCC-grown β-$Ga_2O_3$ remains largely unexplored. In particular, the types of dislocations, their spatial distribution, their correlation with growth morphology, and the long-range lattice strain patterns induced during growth from a cold crucible are unclear. Distinctive defect densities and structural characteristics, compared with those of other growth methods, are expected due to the different heating system configuration employed in OCCC and the higher applicable oxygen partial pressure. Addressing this knowledge gap is crucial for assessing the viability of OCCC as a high-quality substrate technology.

In this work, we performed a detailed structural characterization of OCCC-grown β-$Ga_2O_3$ using a combined approach of synchrotron radiation X-ray topography (SR-XRT) [10] and reticulography [19, 43, 44, 45]. These X-ray-based techniques provide powerful, nondestructive tools for visualizing extended defects with high sensitivity and spatial resolution. In particular, SR-XRT enables quantitative imaging of dislocations, strain fields, and low-angle boundaries over centimeter-scale areas. Reticulography, with its microradian-level sensitivity [19] to lattice tilts and curvature, complements SR-XRT by revealing subtle long-range distortions that cannot be resolved by conventional diffraction methods. Here, reticulography is primarily used to estimate the order of magnitude of local lattice misorientation. A more detailed quantitative analysis, including spatially resolved misorientation mapping, will be reported in a separate study.

## II. EXPERIMENTAL DETAILS

Figure 1(a) schematically illustrates the OCCC setup employed in this study [40]. The β-$Ga_2O_3$ crystal was grown with a pulling direction along the ⟨010⟩ axis using a high-quality seed crystal prepared by the floating-zone method. The seeding, necking, and diameter enlargement processes were carefully optimized. As a result, a crystal with a diameter of approximately 20 mm and a body length exceeding 10 mm was successfully obtained, as shown in Figure 1(b). The growth rates were 6 mm/h before diameter enlargement and 3 mm/h after enlargement, corresponding to regions S and E, respectively. The detailed growth conditions for this crystal have been reported in Ref. 41 and are shown in Fig. 2(c) of that paper.

After growth, the seed portion was removed, and the grown crystal was cleaved along the (100) plane (axial cut), as shown in Figure 1(c). The crystal was then further separated into a central portion and a wing portion, as illustrated in Figure 1(d). The central portion contains the crystal grown beneath the seed before and after diameter enlargement, whereas the wing portion corresponds to the laterally enlarged region. To obtain clear XRT images, the sample surfaces were subjected to chemical–mechanical polishing (CMP) to remove damage introduced during mechanical processing prior to XRT observation.

XRT and reticulography were performed using a SR X-ray source at the BL-3C beamline of the Photon Factory (2.5 GeV storage ring) at the High Energy Accelerator Research Organization (KEK), Tsukuba, Japan. The optical system has been described in detail elsewhere [**46**, **47**, **48**]; therefore, only the essential components are briefly summarized here for completeness. Figure 2 schematically illustrates the experimental setup used for XRT and reticulography. Monochromatic SR X-rays with an effective divergence of approximately $7.3 \times 10^{-6}$ rad were incident on the crystal surface at an angle of 5°. The X-ray beam size was approximately 16 mm (horizontal) × 7 mm (vertical). At an incident angle of 5°, the irradiated area was approximately 16 mm × 80 mm, which was sufficient to cover the entire crystal. The X-ray wavelength was selected according to the reciprocal lattice vector used for diffraction (the $\boldsymbol{g}$-vector). The X-ray wavelength was 0.166 nm for the $\boldsymbol{g}$ = 710 reflection and 0.148 nm for the $\boldsymbol{g}$ = $10\ 0\overline{3}$ reflection. The corresponding X-ray penetration depth in β-$Ga_2O_3$ was estimated to be 10~20 μm. The diffracted X-rays were recorded by two-dimensional detectors to obtain XRT images, using either a digital CCD detector or an analog X-ray film (Agfa Structurix D2). When a CCD detector was employed, a sequence of XRT images could be continuously acquired while varying the incident angle, a technique referred to as ω-rocking X-ray diffraction imaging (XRDI). XRT images (as shown later in Figures 6–9) were recorded on X-ray films using a single exposure of several seconds under fixed diffraction conditions.

For reticulography, a fine X-ray-absorbing mesh with a period of r = 100 μm and a 50 μm–50 μm aperture-to-bar ratio was placed between the sample and the detector [**19**]. The X-rays diffracted from the crystal were divided by the mesh into an array of distinguishable microbeams, each propagating in a slightly different direction due to spatial inhomogeneities in the crystal orientation. Upon reaching the recording medium, each microbeam produced an individual XRT image corresponding to the mesh aperture.

The relative displacement among these microbeam images enables quantitative determination of the local misorientation of the crystal region responsible for diffracting each microbeam [43, 44].

## III. RESULTS AND DISCUSSION

Figure 3(a) shows an optical image of the central portion of the crystal. Beneath the seed, the crystal was initially grown to a length of 27 mm with a stable diameter of approximately 7 mm. Subsequently, the diameter was increased, and the shoulder region can be clearly identified. To evaluate the crystallinity and dislocation density at different stages of crystal growth, the XRT observations were divided into three regions: S (start), M (middle), and E (end), corresponding to crystal growth before, during, and after the diameter enlargement. These regions are indicated by the red, green, and blue frames, respectively.

Due to the curvature of the lattice planes, it was difficult to obtain an XRT image over the entire area of interest, as the Bragg condition was satisfied only locally. Therefore, for each region, a sequence of XRDI images was acquired by rocking the ω angle, and a composite image was then constructed by selecting the maximum diffraction intensity at each pixel of the two-dimensional detector. The ω-rocking range depends on the degree of lattice-plane curvature; in other words, the more uniform the crystal orientation, the narrower the required ω-rocking range. It is noted that the observed lattice plane curvature reflects the combined effect of macroscopic lattice bending and local lattice distortions associated with structural defects such as dislocations and stacking faults. In the present study, these contributions are not quantitatively separated, and the contrast observed in the XRT images represents the total local lattice misorientation. In this study, the ω-rocking ranges were 0.1°, 0.1°, and 0.34° for regions S, M, and E, respectively. The resulting ω-rocking XRDI images constitute a so-called hyperspectral dataset, in which conventional X-ray ω-rocking curves are recorded for every pixel. Details of this technique have been reported elsewhere [49]. The diffraction vector ***g*** = 710 was used. Figure 3(h) shows the (710) plane.

Figure 3(b)–(d) show the maximum-intensity images constructed from sequences of XRDI data for regions S, M, and E, respectively. Movies of the ω-dependent XRT images for the three regions are provided as Supplementary Movies 1–3, respectively. A ring-shaped marker was placed near the shoulder to indicate its position. As shown in Figure 3(b), the region S grown directly beneath the seed exhibited a uniform diffraction intensity, indicating high crystallinity. After an initial growth of approximately 10 mm, a significant disturbance in crystal orientation occurred, as evidenced by the appearance of dark regions in the peripheral areas relative to the axial center. These dark regions in the XRT images indicate the presence of lattice misorientation. However, this does not necessarily imply that the misorientation exceeded the ω-rocking range; it may instead result from a deviation of the reciprocal lattice vector from the plane of scattering [19, 49, 50, 51, 52]. Consequently, the X-rays diffracted from this region of the crystal propagate out of the plane defined by the incident X-rays and the detector and therefore do not reach the detector. Subsequent reticulography and XRT observations using ***g*** = $10\ 0\bar{3}$ ($h$ = 10, $k$ = 0, $l$ = −3) confirmed that this lattice misorientation was caused by the formation of a domain boundary that is detectable with ***g*** = 710 but not with ***g*** = $10\ 0\bar{3}$, as will be explained in detail later.

As shown in Figure 3(c), during diameter enlargement, the region directly beneath the seed maintained the same crystal orientation, whereas the laterally enlarged region did not. Although separate XRT observations confirmed that the enlarged region is also a single crystal of β-$Ga_2O_3$ with similarly high crystalline quality, it was misoriented with respect to the central region along the growth axis. This misorientation is considered to arise because diameter enlargement is primarily achieved through lateral crystal growth along directions orthogonal to ⟨010⟩; in the absence of a seed above the enlarged region, maintaining the same crystal orientation as the seed becomes difficult. Similar issues, such as the formation of twin boundaries near the shoulder, have also been reported in the EFG technique [11]. Therefore, achieving diameter enlargement while preserving a uniform crystal orientation during OCCC growth remains an important subject for future research.

Further down toward the end region, most areas are out of the Bragg condition for $\boldsymbol{g}$ = 710, except for those near the growth axis, and therefore appear dark in the XRT image (Figure 3(d)). A fracture-like feature is visible in the lower-left part of the image; however, it does not correspond to an actual physical fracture of the crystal. Rather, it results from a locally misoriented domain that diffracts X-rays in a direction slightly deviated from the ideal 2θ angle for $\boldsymbol{g}$ = 710. Similar to the case shown in Figure 3(c), subsequent XRT measurements recorded using $\boldsymbol{g}$ = $10\,0\bar{3}$ confirmed that most of the misorientation in region E is a twist-type misorientation associated with the growth axis. In other words, the bright regions in the XRT image that satisfy the Bragg condition for $\boldsymbol{g}$ = 710 and those that do not, share the same ⟨010⟩ direction as the growth direction but are rotated relative to each other about the $\boldsymbol{b}$-axis.

To evaluate the crystallinity of the OCCC crystal in terms comparable to conventional laboratory-based XRD measurements, we reconstructed ω-rocking curves by extracting 1 mm × 3 mm regions from the stack of XRDI images [49], as indicated by the yellow frames in Figure 3(b)–(d). This region size is comparable to the typical irradiation area used in laboratory XRD instruments. The integrated intensity of all pixels within each selected region was then plotted as a function of ω to obtain the corresponding rocking curve. As shown in Figure 3(e)–(g), region S exhibits a FWHM of 26 arcsec, whereas region M shows a double-peak structure, with each peak having a FWHM of less than 50 arcsec. This double-peak structure is consistent with the twist-type misorientation discussed above. When the selected area does not contain multiple domains, the rocking curve exhibits a sharp single peak, as observed for region E, with a FWHM of 37 arcsec. These values are only marginally broader than those reported for commercial EFG crystals [46], indicating the strong potential of OCCC as a bulk crystal growth technique. It should be noted that the double-peak structure reflects only low-angle misorientation components within the accessible ω-rocking range, while domains with larger twist misorientation fall outside this range and do not contribute to the observed rocking curve.

To elucidate how and in which direction the disturbance in crystal orientation occurred during diameter enlargement, we acquired another set of XRDI images using the ***g***-vector $10\ 0\bar{3}$. Figure 4(a)–(c) show the maximum-intensity XRT images of the three regions obtained with ***g*** = $10\ 0\bar{3}$. These images can be regarded as counterparts to Figure 3(b)–(d), differing only in the ***g***-vector used. Movies of the ω-dependent XRT images for the three regions are provided as Supplementary Movies 4–6, respectively. The ω ranges are 0.2°, 0.7°, and 1.0°, respectively. Compared with ***g*** = 710, the ***g***-vector $10\ 0\bar{3}$ is an *h*0*l*-type reflection containing only ***a***-axis and ***c***-axis components. As shown in Figure 4(d), the ***g***-vector $10\ 0\bar{3}$ corresponds to the lattice planes (pink plane) that contain the ***b***-axis. With this ***g***-vector, if the ***b***-axes of different parts of the crystal are tilted relative to one another, their $10\ 0\bar{3}$ diffractions deviate such that, regardless of ω rotation, the diffracted X-rays cannot enter the detector. In other words, a tilt of the ***b***-axis leads to a deviation of the diffraction out of the plane of scattering.

As shown in Figure 4(a), Region S (the portion beneath the seed) exhibits high crystallinity and uniform crystal orientation, similar to that observed in Figure 3. Disturbance of the crystal orientation occurs near the shoulder, as indicated by the appearance of dark regions in Figure 4(b) and (c). However, unlike the case of $\boldsymbol{g}$ = 710, in which the enlarged region was almost completely dark, the 10 0$\overline{3}$ images show bright contrast beneath the shoulder. This observation indicates that the enlarged region shares the same ***b***-axis orientation as the central region but exhibits a twist misorientation with the ***b***-axis as the twist axis. During diameter enlargement, crystal growth proceeds laterally in directions perpendicular to this axis, where the constraint imposed by the seed crystal becomes weaker. Under such conditions, the growth direction tends to be maintained, while small rotational deviations around the ⟨010⟩ axis can occur, resulting in twist-type misorientation. This behavior may also be influenced by the anisotropic nature of the monoclinic β-$Ga_2O_3$ structure, in which the ⟨010⟩ direction is crystallographically distinct. As a result, misorientation may preferentially develop as a rotation around this axis rather than as a tilt. A quantitative analysis of this mechanism is beyond the scope of the present study. Similar twist misorientations with domain boundaries extending along the ⟨010⟩ direction have been reported in EFG crystals. [19].

X-ray reticulography is a powerful technique for investigating lattice misorientation with high angular sensitivity on the order of $10^{-6}$ arcsec. Figures 5(a) and (c) show the reticulography images of region M taken with $\boldsymbol{g}$ = 710 and $\boldsymbol{g}$ = 10 0$\overline{3}$, respectively. Figures 5(b) and (d) are magnified images of the same area indicated by the red frame. Compared with typical XRT images of region M shown in Figures 3 and 4, where the image brightness or contrast changes only slightly when the lattice misorientation is small, reticulography enables us to extract the characteristics of the misorientation in terms of twist or tilt with respect to a particular axis, as well as the magnitude of the misorientation. This is possible because each bright spot in the image originates from a specific position in the crystal with a certain crystal orientation [19, 43]. For example, in the upper part of region M, an obvious onset of misorientation occurs near the position indicated by the red arrow. This misorientation then splits into two boundaries extending roughly along the growth direction. One of these boundaries is indicated by the dashed yellow line. The left side of this boundary has a similar crystal orientation to region S and to the seed, as the spot matrix is well aligned with that observed in region S. In contrast, the right side of the boundary is out of contrast at $\boldsymbol{g}$ = 710. When this boundary is observed with $\boldsymbol{g}$ = 10 0$\bar{3}$, it becomes clear from Figure 5(c) that the spot matrix on the right side is rotated with respect to that on the left side. This can be recognized more clearly in Figure 5(e).

The second boundary, which also originates from the red arrow, curves toward the left side and appears as a white line in Figure 5(a) and as a black line in Figure 5(c). Compared with the yellow boundary, the degree of misorientation across this boundary is smaller, since both sides of the boundary remain visible for both $\boldsymbol{g}$-vectors. Zooming into the details of the spot matrix in the area marked by the red frame, it is clear that the spots are well aligned along the $\boldsymbol{b}$-axis, and the two sides of the boundary do not show any significant rotation of the spot matrix for either $\boldsymbol{g}$ = 710 or $\boldsymbol{g}$ = 10 0$\bar{3}$. In particular, the boundary becomes almost unrecognizable at $\boldsymbol{g}$ = 10 0$\bar{3}$, as there is no obvious rotation or displacement of the spot matrix, indicating an extremely small misorientation on the order of $10^{-6}$ arcsec [19]. Nevertheless, the presence of this misorientation can still be detected when the coordinates of the spots are carefully extracted and analyzed, as shown in Figure 5(f). Domain boundaries with misorientations approximately one order of magnitude larger (~$10^{-5}$ arcsec) are typically found in regions after the diameter enlargement. A quantitative analysis of misorientation using reticulography in combination with ray-tracing simulations will be reported in a separate paper.

Since the performance and reliability of power devices are strongly influenced by dislocations in the semiconductor substrate, understanding the dislocation characteristics in OCCC-grown crystals is essential. In this section, we analyze the density and character of dislocations in the OCCC crystal using XRT. Figure 6(b) shows an XRT image of an area selected from region S, indicated by the green frame in Figure 6(a). The image was obtained using ***g*** = 710 at a fixed ω angle with an X-ray film. Because X-ray films produce a negative image, darker regions correspond to areas of stronger X-ray exposure. Except for a small region at the top of the image, the entire area of approximately 7 mm × 27 mm satisfies the Bragg condition at the same ω angle. This observation indicates a high crystalline quality, with highly aligned lattice planes and very small lattice curvature across the measured region. It should be noted that such a wide area satisfying the Bragg condition was not observed during the early stage of OCCC development (~year 2020). At that time, XRT images typically exhibited very narrow linear features, indicating that the crystal orientation varied locally. As a result, the Bragg condition was satisfied only within small regions, reflecting a significant local misorientation of the lattice. Two representative regions, P1 and P2, were selected for detailed dislocation analysis, as shown in Figures 6(c) and 6(d). In the region close to the seed crystal, corresponding to the early growth stage, two types of dislocations are observed. The first type consists of straight dislocations extending along the ⟨010⟩ direction, which appear as vertical lines in Figure 6(c). The second type corresponds to curved dislocations lying on the (100) plane, which is parallel to the sample surface. Based on the XRT contrast of dislocations observed at ***g*** = 710 and ***g*** = $10\ 0\bar{3}$ (as shown later in Figure 8), the ⟨010⟩-oriented dislocations are attributed to screw-type dislocations with both their Burgers vector and line direction parallel to ⟨010⟩. This assignment is reasonable because the Burgers vector b = ⟨010⟩ corresponds to the smallest lattice translation vector in monoclinic β-$Ga_2O_3$, with a magnitude of 0.304 nm. Dislocations with the smallest Burgers vector are energetically favorable and are therefore commonly observed in β-$Ga_2O_3$ crystals. Indeed, similar ⟨010⟩-type screw dislocations have also been reported as the dominant dislocation type in EFG-grown crystals [**18**, **53**].

The character of the curved dislocations is less clear. Their visibility under $\boldsymbol{g}$ = 710 suggests that their Burgers vectors contain either $\langle h00\rangle$, $\langle 0k0\rangle$, or combinations of these components. Because the diffraction vector $\boldsymbol{g}$ = 710 is orthogonal to $\langle 00l\rangle$-type Burgers vectors, the presence of a $\boldsymbol{c}$-axis component cannot be determined from this reflection alone based on the $\boldsymbol{g}\cdot\boldsymbol{b}$ invisibility criterion. However, additional geometric considerations provide further constraints. These dislocations exhibit a curved morphology and lie on the (100) plane, suggesting that they may belong to the ⟨010⟩{100} slip system. If this interpretation is correct, the Burgers vector cannot contain a $\langle h00\rangle$ component (i.e., an $\boldsymbol{a}$-axis component), because such a component would cause the Burgers vector to deviate from the (100) plane and would therefore prevent glide on this plane. On the other hand, the presence of a $\langle 00l\rangle$ component (i.e., a $\boldsymbol{c}$-axis component) does not contradict the assumption that (100) is the slip plane. In this case, however, the slip system would correspond to ⟨001⟩{100}, as reported in previous studies [16, 18]. In the present study, the limited number of g-vectors prevents an unambiguous analysis of the Burgers vector, and the interpretation remains tentative. Further diffraction analyses using additional $\boldsymbol{g}$-vectors will be required to unambiguously determine the Burgers vectors of these dislocations.

A comparison between regions P1 and P2 reveals a significant difference in dislocation morphology. In P2, the majority of dislocations appear as vertical lines aligned along the ⟨010⟩ direction. Once ⟨010⟩-oriented screw dislocations are generated, they tend to propagate along the growth direction during the pulling process. This behavior can be understood from the crystallographic geometry: these dislocations possess the smallest Burgers vector and their line direction is nearly parallel to the growth direction, making them energetically favorable for propagation along the advancing growth front. In addition to the vertical dislocations, spot-like contrasts are observed in region P2, whereas they are almost absent in P1. These spot-like contrasts correspond to dislocations whose line directions are not confined to the (100) plane, indicating the presence of additional dislocation configurations in the later growth stage. The dislocation densities were estimated assuming an X-ray penetration depth of 20 μm. In region P1, the densities corresponding to the vertical dislocations and the curved dislocations are estimated to be $1.0 \times 10^5$ cm$^{-2}$ and $1.0 \times 10^4$ cm$^{-2}$, respectively. In region P2, the densities corresponding to the vertical dislocations and the spot-like contrasts are approximately $1.0 \times 10^5$ cm$^{-2}$ and $1.6 \times 10^4$ cm$^{-2}$, respectively. The curved morphology of these dislocations may be associated with dislocation climb processes. In addition, the spot-like contrasts observed in later growth stages could be related to changes in dislocation line direction, possibly involving cross-slip or other three-dimensional dislocation configurations. However, the present data do not allow a definitive identification of these mechanisms.

Further downward toward region M, as shown in Figure 7, the dislocation characteristics do not change significantly compared with those observed in the lower part of region S. Vertical line contrasts remain the dominant features in the XRT images and are attributed to ⟨010⟩-oriented screw dislocations. Near the diameter enlargement region, however, ripple-like contrast variations begin to appear. The presence of these variations suggests temporal fluctuations in local strain or other inhomogeneities in the crystal, such as impurity incorporation during this stage of crystal growth [45, 54, 55, 56, 57, 58]. Their appearance in the present OCCC crystal therefore indicates that the growth process experiences comparable transient variations in the growth environment during diameter enlargement.

The assignment of the vertical line contrasts to ⟨010⟩-oriented screw dislocations was verified by comparing XRT images obtained from the same areas using $\boldsymbol{g}$ = 710 and $\boldsymbol{g}$ = 10 $0\bar{3}$. Figure 8 shows two representative regions used for this comparison. In region P3, the vertical lines are clearly visible when $\boldsymbol{g}$ = 710 is used (Figure 8(c)), whereas they completely disappear in the image taken with $\boldsymbol{g}$ = $10\ 0\bar{3}$ (Figure 8(b)). This disappearance indicates that $\boldsymbol{g}\cdot\boldsymbol{b}$ = 0 for $\boldsymbol{g}$ = $10\ 0\bar{3}$, confirming that these dislocations possess a ⟨010⟩ Burgers vector, consistent with ⟨010⟩-oriented screw dislocations. In contrast, dislocation bundles inclined away from the ⟨010⟩ direction are clearly visible in the $\boldsymbol{g}$ = $10\ 0\bar{3}$ image but are only weakly visible when $\boldsymbol{g}$ = 710 is used. This observation suggests that these dislocations possess a $\boldsymbol{c}$-axis Burgers vector component. The weak contrast observed in the $\boldsymbol{g}$ = 710 image is likely a residual contrast, arising from the fact that these dislocations are not pure screw dislocations but contain mixed character [59]. It is worth noting that the weak residual contrast is likely due not only to the mixed character of the dislocations but also to the limitations of the g·b invisibility criterion in the monoclinic crystal system, where $\boldsymbol{g}\cdot\boldsymbol{b}$ = 0 may not result in complete extinction of contrast. A similar behavior is observed in region P4. The vertical lines are clearly visible in Figure 8(e) but disappear in Figure 8(d) when the diffraction vector changes, again indicating that these dislocations satisfy the $\boldsymbol{g}\cdot\boldsymbol{b}$ = 0 invisibility condition for the latter reflection.

If the curved dislocations indeed possess a $\boldsymbol{c}$-axis Burgers vector, their local character changes depending on the orientation of the dislocation line. For example, the segments indicated by the green arrows exhibit an edge-type character, whereas segments that bend toward the [001] direction, indicated by the yellow arrows, exhibit a screw-type character. This interpretation explains the observed contrast behavior: the green-arrow segments show weak residual contrast in Figure 8(e), while the yellow-arrow segments completely lose their contrast due to the $\boldsymbol{g}\cdot\boldsymbol{b}$ invisibility condition.

Finally, the wing portion of the OCCC crystal was analyzed. The XRT image shown in Figure 9(a) reveals a high density of ripple-like contrast variations this region. XRD analyses were performed for five areas, each with dimensions of 1 mm in height and 3 mm in width, selected from the wing portion as indicated by the yellow frames. The XRD curves were generated using the same method described previously for Figure 3. The measured FWHM values range from 33 to 60 arcsec. Areas 1 and 5 exhibit a single-peak profile, whereas areas 2, 3, and 4 show clear peak shoulders, indicating a disturbance in the crystal orientation during this stage of growth. A high-resolution XRT image corresponding to the green frame is shown in Figure 9(c). A dislocation bundle is observed on the left side of the region, which appears to propagate downward, with its onset located near the wing shoulder. The dislocations are aligned either vertically or slightly inclined from the ⟨010⟩ direction, with a relatively high density of approximately $10^6$ $cm^{-2}$. The curved, parallel lines observed near the left edge of the image are attributed to imaging artifacts, likely arising from edge-related diffraction or secondary scattering effects, and do not represent intrinsic crystal defects. Overall, the crystallinity in the wing portion is inferior to that in regions S and M beneath the seed. A movie of the ω-dependent XRT images for the wing portion are provided as Supplementary Movie 7.

## IV. CONCLUSIONS

In this study, the structural properties and defect structure of a β-$Ga_2O_3$ single crystal grown by the OCCC method were systematically investigated using SR-XRT and X-ray reticulography. These techniques enabled nondestructive visualization of lattice distortions, domain boundaries, and dislocations over large areas with high spatial and angular sensitivity.

The region grown directly beneath the seed exhibited highly uniform diffraction contrast and excellent crystalline quality. Rocking curves reconstructed from ω-rocking X-ray diffraction imaging showed a FWHM of approximately 26 arcsec, which is comparable to values reported for commercial melt-grown β-$Ga_2O_3$ substrates. However, during diameter enlargement, a twist-type lattice misorientation developed between the central region and the laterally expanded portion of the crystal. Reticulography measurements revealed that this misorientation originates near the shoulder region and propagates downward along domain boundaries roughly parallel to the ⟨010⟩ growth direction. The magnitude of the misorientation across these boundaries was estimated to be on the order of $10^{-6}$–$10^{-5}$ rad, which should be regarded as the total local lattice misorientation, including contributions from both lattice plane curvature and defect-related distortions.

Detailed dislocation analysis revealed that the dominant defect type in the central region is the ⟨010⟩-oriented screw dislocation. The assignment of these dislocations was confirmed using the $\boldsymbol{g}\cdot\boldsymbol{b}$ invisibility criterion by comparing XRT images obtained with different diffraction vectors. The density of these screw dislocations was estimated to be approximately $10^5$ cm$^{-2}$. Curved dislocations lying on the (100) plane were also observed near the seed region and are likely associated with slip systems involving ⟨010⟩{100} or ⟨001⟩{100}. As growth proceeds toward the diameter enlargement region, ripple-like contrast variations appear, indicating temporal fluctuations in growth conditions. In the wing region formed during lateral crystal expansion, the crystalline quality deteriorates relative to the central region. This region exhibits higher dislocation densities on the order of $10^6$ cm$^{-2}$ and broader rocking curve widths (33–60 arcsec). These results suggest that diameter enlargement is a critical stage for defect generation in OCCC growth.

Overall, the present work provides a detailed characterization of lattice misorientation, domain boundaries, and dislocations in OCCC-grown β-$Ga_2O_3$ crystals. While relatively high crystalline quality was achieved in the region beneath the seed, the results also indicate that defect formation during diameter enlargement remains a critical issue. These findings highlight both the potential and the current limitations of the OCCC growth method, which is still at an early stage of development. Further optimization of growth conditions will be necessary to improve crystal quality and achieve more consistent structural properties.

## Supplementary Material

See supplementary movie 1–3 for the ω-dependent XRT images for regions S, M, and E, respectively, taken using $\boldsymbol{g}$ = 710.

See supplementary movie 4–6 for the ω-dependent XRT images for regions S, M, and E, respectively, taken using $\boldsymbol{g}$ = 10 0$\bar{3}$.

See supplementary movie 7 for the ω-dependent XRT images for the wing portion taken using $\boldsymbol{g}$ = 710.

## Acknowledgments

This study was financially supported by (1) the MEXT Program for Creation of Innovative Core Technology for Power Electronics (Grant Number JPJ009777), Japan, (2) JSPS Grant-in-Aid for Scientific Research (A) (Grant Number 25H00725), Japan, (3) JSPS KAKENHI Grant No. 20K05355, 23H01872, and 23K17356, Japan, (4) the Murata Science Foundation, Japan, (5) the Sumitomo Foundation, Japan, (6) the Kurata Grants by the Hitachi Global Foundation, (7) the Iketani Science and Technology Foundation, and (8) the Iwatani Naoji Foundation. The synchrotron XRT observations were performed at KEK-PF under proposal Nos. 2018G501, 2020G585, 2022G503, and 2024G520.

## AUTHOR DECLARATIONS

### Declaration of competing interests

The authors declare that they have no known competing financial interests or personal relationships that could have appeared to influence the work reported in this paper.

### Generative AI

Not used in the manuscript preparation process.

### Author contributions

Yongzhao Yao: Conceptualization (lead); Data curation (lead); Formal analysis (lead); Investigation (lead); Methodology (lead); Project administration (lead); Writing - original draft (lead); Writing - review & editing (lead);

Koki Mizuno: Data curation (equal); Writing - review & editing (equal);

Kazuki Ohnishi: Data curation (equal); Writing - review & editing (equal);

Yukari Ishikawa: Resources (equal); Writing - review & editing (equal);

Masanori Kitahara: Investigation (equal); Resources (equal); Writing - review & editing (equal);

Taketoshi Tomida: Investigation (equal); Resources (equal); Writing - review & editing (equal);

Rikito Murakami: Investigation (equal); Resources (equal); Writing - review & editing (equal);

Vladimir Kochurikhin: Investigation (equal); Resources (equal); Writing - review & editing (equal);

Liudmila Gushchina: Investigation (equal); Resources (equal); Writing - review & editing (equal);

Kei Kamada: Investigation (equal); Resources (equal); Writing - review & editing (equal);

Koichi Kakimoto: Investigation (equal); Resources (equal); Writing - review & editing (equal);

Akira Yoshikawa: Funding acquisition (lead); Investigation (equal); Project administration (lead); Resources (equal); Writing - review & editing (equal);

All authors have approved the manuscript.

**Data Availability**

Raw data were generated at the synchrotron facilities KEK-PF. The data that support the findings of this study are available within the article and its supplementary material.

## REFERENCES

[1] S. Pearton, J. Yang, P. Cary, F. Ren, J. Kim, M. Tadjer, and M. Mastro, A review of $Ga_2O_3$ materials, processing, and devices, Appl. Phys. Rev. 5, 011301 (2018). [DOI:10.1063/1.5006941]

[2] S. Pearton, F. Ren, M. Tadjer, and J. Kim, Perspective: $Ga_2O_3$ for ultra-high power rectifiers and MOSFETS, J. Appl. Phys. 124, 220901 (2018). [DOI:10.1063/1.5062841]

[3] M. Higashiwaki, β-$Ga_2O_3$ material properties, growth technologies, and devices: a review, AAPPS Bull. 32, 3 (2022). [DOI:10.1007/s43673-021-00033-0]

[4] K. Sasaki, Prospects for β-$Ga_2O_3$: now and into the future, Appl. Phys. Express 17, 090101-1 (2024). [DOI:10.35848/1882-0786/ad6b73]

[5] O. Ueda, M. Kasu, and H. Yamaguchi, Structural characterization of defects in EFG- and HVPE-grown β-$Ga_2O_3$ crystals, Jpn. J. Appl. Phys. 61, 050101 (2022). [DOI:10.35848/1347-4065/ac4b6b]

[6] M. Kasu, K. Hanada, T. Moribayashi, A. Hashiguchi, T. Oshima, T. Oishi, K. Koshi, K. Sasaki, A. Kuramata, and O. Ueda, Relationship between crystal defects and leakage current in β-$Ga_2O_3$ Schottky barrier diodes, Jpn. J. Appl. Phys. 55, 1202BB (2016). [DOI:10.7567/JJAP.55.1202BB]

[7] M. Kasu, T. Oshima, K. Hanada, T. Moribayashi, A. Hashiguchi, T. Oishi, K. Koshi, K. Sasaki, A. Kuramata, and O. Ueda, Crystal defects observed by the etch-pit method and their effects on Schottky-barrier-diode characteristics on (-201) β-$Ga_2O_3$, Jpn. J. Appl. Phys. 56, 091101 (2017). [DOI:10.7567/JJAP.56.091101]

[8] S. Sdoeung, K. Sasaki, A. Kuramata, and M. Kasu, Identification of dislocation responsible for leakage current in halide vapor phase epitaxial (001) β-$Ga_2O_3$ Schottky barrier diodes investigated via ultrahigh-sensitive emission microscopy and synchrotron X-ray topography, Appl. Phys. Express 15, 111001 (2022). [DOI:10.35848/1882-0786/ac9726]

[9] S. Sdoeung, K. Sasaki, S. Masuya, K. Kawasaki, J. Hirabayashi, A. Kuramata, and M. Kasu, Stacking faults: Origin of leakage current in halide vapor phase epitaxial (001) β-$Ga_2O_3$ Schottky barrier diodes, Appl. Phys. Lett. 118, 172106 (2021). [DOI:10.1063/5.0049761]

[10] Y. Yao, D. Wakimoto, H. Miyamoto, K. Sasaki, A. Kuramata, K. Hirano, Y. Sugawara, and Y. Ishikawa, X-ray topographic observation of dislocations in β-$Ga_2O_3$ Schottky barrier diodes and their glide and multiplication under reverse bias, Scr. Mater. 226, 115216 (2023). [DOI:10.1016/j.scriptamat.2022.115216]

[11] A. Kuramata, K. Koshi, S. Watanabe, Y. Yamaoka, T. Masui, and S. Yamakoshi, High-quality β-$Ga_2O_3$ single crystals grown by edge-defined film-fed growth, Jpn. J. Appl. Phys. 55, 1202A2 (2016). [DOI:10.7567/JJAP.55.1202A2]

[12] S. Reese, T. Remo, J. Green, and A. Zakutayev, How much will gallium oxide power electronics cost?, Joule 3, 903 (2019). [DOI:10.1016/j.joule.2019.01.011]

[13] J. Jesenovec, M. Weber, C. Pansegrau, M. McCluskey, K. Lynn, and J. McCloy, Gallium vacancy formation in oxygen annealed β-$Ga_2O_3$, J. Appl. Phys. 129, 245701 (2021). [DOI:10.1063/5.0053325]

[14] M. McCluskey, Point defects in $Ga_2O_3$, J. Appl. Phys. 127, 101101 (2020). [DOI:10.1063/1.5142195]

[15] Y. Yao, K. Hirano, Y. Sugawara, K. Sasaki, A. Kuramata, and Y. Ishikawa, Observation of dislocations in thick β-$Ga_2O_3$ single-crystal substrates using Borrmann effect synchrotron x-ray topography, APL Mater. 10, 051101 (2022). [DOI:10.1063/5.0088701]

[16] Y. Yao, Y. Tsusaka, K. Sasaki, A. Kuramata, Y. Sugawara, and Y. Ishikawa, Large-area total-thickness imaging and Burgers vector analysis of dislocations in β-$Ga_2O_3$ using bright-field x-ray topography based on anomalous transmission, Appl. Phys. Lett. 121, 012105 (2022). [DOI:10.1063/5.0098942]

[17] Y. Yao, Y. Tsusaka, K. Hirano, K. Sasaki, A. Kuramata, Y. Sugawara, and Y. Ishikawa, Three-dimensional distribution and propagation of dislocations in β-$Ga_2O_3$ revealed by Borrmann effect X-ray topography, J. Appl. Phys. 134, 155104 (2023). [DOI:10.1063/5.0169526]

[18] Y. Yao, Y. Sugawara, and Y. Ishikawa, Identification of Burgers vectors of dislocations in monoclinic β-$Ga_2O_3$ via synchrotron x-ray topography, J. Appl. Phys. 127, 205110 (2020). [DOI:10.1063/5.0007229]

[19] Y. Yao, K. Hirano, K. Sasaki, A. Kuramata, Y. Sugawara, and Y. Ishikawa, Lattice misorientation at domain boundaries in β-$Ga_2O_3$ single-crystal substrates observed via synchrotron radiation X-ray diffraction imaging and X-ray reticulography, J. Am. Ceram. Soc. 106, 5487 (2023). [DOI:10.1111/jace.19156]

[20] O. Ueda, N. Ikenaga, K. Koshi, K. Iizuka, A. Kuramata, K. Hanada, T. Moribayashi, S. Yamakoshi, and M. Kasu, Structural evaluation of defects in β-$Ga_2O_3$ single crystals grown by edge-defined film-fed growth process, Jpn. J. Appl. Phys. 55, 1202BD (2016). [DOI:10.7567/JJAP.55.1202BD]

[21] K. Nakai, T. Nagai, K. Noami, and T. Futagi, Characterization of defects in β-$Ga_2O_3$ single crystals, Jpn. J. Appl. Phys. 54, 051103 (2015). [DOI:10.7567/JJAP.54.051103]

[22] K. Ogawa, N. Ogawa, R. Kosaka, T. Isshiki, Y. Yao, and Y. Ishikawa, Three-dimensional observation of internal defects in a β-$Ga_2O_3$ (001) wafer using the FIB–SEM serial sectioning method, J. Electron. Mater. 49, 5190 (2020). [DOI:10.1007/s11664-020-08313-5]

[23] K. Ogawa, N. Ogawa, R. Kosaka, T. Isshiki, T. Aiso, M. Iyoki, Y. Yao, and Y. Ishikawa, AFM observation of etch-pit shapes on β-$Ga_2O_3$ (001) surface formed by molten alkali etching, Mater. Sci. Forum 1004, 512 (2020). [DOI:10.4028/www.scientific.net/MSF.1004.512]

[24] K. Ogawa, K. Kobayashi, N. Hasuike, and T. Isshiki, Crystal structure analysis of stacking faults through scanning transmission electron microscopy of β-$Ga_2O_3$ (001) layer grown via halide vapor phase epitaxy, J. Vac. Sci. Technol. A 40, 032701 (2022). [DOI:10.1116/6.0001799]

[25] S. Isaji, I. Maeda, N. Ogawa, R. Kosaka, N. Hasuike, T. Isshiki, K. Kobayashi, Y. Yao, and Y. Ishikawa, Relationship between propagation angle of dislocations in β-$Ga_2O_3$ (001) bulk wafers and their etch pit shapes, J. Electron. Mater. 52, 5093 (2023). [DOI:10.1007/s11664-023-10363-4]

[26] Z. Galazka, Growth of bulk β-$Ga_2O_3$ single crystals by the Czochralski method, J. Appl. Phys. 131, 031103 (2022). [DOI:10.1063/5.0076962]

[27] Z. Galazka, β-$Ga_2O_3$ for wide-bandgap electronics and optoelectronics, Semicond. Sci. Technol. 33, 113001 (2018). [DOI:10.1088/1361-6641/aadf78]

[28] Z. Galazka, S. Ganschow, P. Seyidov, K. Irmscher, M. Pietsch, T. Chou, S. Bin Anooz, R. Grueneberg, A. Popp, A. Dittmar, A. Kwasniewski, M. Suendermann, D. Klimm, T. Straubinger, et al., Two inch diameter, highly conducting bulk β-$Ga_2O_3$ single crystals grown by the Czochralski method, Appl. Phys. Lett. 120, 152101 (2022). [DOI:10.1063/5.0086996]

[29] Z. Galazka, S. Ganschow, K. Irmscher, D. Klimm, M. Albrecht, R. Schewski, M. Pietsch, T. Schulz, A. Dittmar, A. Kwasniewski, R. Grueneberg, S. Anooz, A. Popp, U. Juda, et al., Bulk single crystals of β-$Ga_2O_3$ and Ga-based spinels as ultra-wide bandgap transparent semiconducting oxides, Prog. Cryst. Growth Charact. Mater. 67, 100511 (2021). [DOI:10.1016/j.pcrysgrow.2020.100511]

[30] Z. Galazka, K. Irmscher, R. Uecker, R. Bertram, M. Pietsch, A. Kwasniewski, M. Naumann, T. Schulz, R. Schewski, D. Klimm, and M. Bickermann, On the bulk β-$Ga_2O_3$ single crystals

grown by the Czochralski method, J. Cryst. Growth 404, 184 (2014). [DOI:10.1016/j.jcrysgro.2014.07.021]

[31] K. Hoshikawa, T. Kobayashi, E. Ohba, and T. Kobayashi, 50 mm diameter Sn-doped (001) β-$Ga_2O_3$ crystal growth using the vertical Bridgeman technique in ambient air, J. Cryst. Growth 546, 125778 (2020). [DOI:10.1016/j.jcrysgro.2020.125778]

[32] K. Hoshikawa, E. Ohba, T. Kobayashi, J. Yanagisawa, C. Miyagawa, and Y. Nakamura, Growth of β-$Ga_2O_3$ single crystals using vertical Bridgman method in ambient air, J. Cryst. Growth 447, 36 (2016). [DOI:10.1016/j.jcrysgro.2016.04.022]

[33] K. Hoshikawa, T. Kobayashi, Y. Matsuki, E. Ohba, and T. Kobayashi, 2-inch diameter (1 0 0) β-$Ga_2O_3$ crystal growth by the vertical Bridgman technique in a resistance heating furnace in ambient air, J. Cryst. Growth 545, 125724 (2020). [DOI:10.1016/j.jcrysgro.2020.125724]

[34] E. Ohba, T. Kobayashi, T. Taishi, and K. Hoshikawa, Growth of (100), (010) and (001) β-$Ga_2O_3$ single crystals by vertical Bridgman method, J. Cryst. Growth 556, 125990 (2021). [DOI:10.1016/j.jcrysgro.2020.125990]

[35] E. Ohba, T. Kobayashi, M. Kado, and K. Hoshikawa, Defect characterization of β-$Ga_2O_3$ single crystals grown by vertical Bridgman method, Jpn. J. Appl. Phys. 55, 1202BF (2016). [DOI:10.7567/JJAP.55.1202BF]

[36] T. Igarashi, Y. Ueda, K. Koshi, R. Sakaguchi, S. Watanabe, S. Yamakoshi, and A. Kuramata, Growth of 6 inch diameter β-$Ga_2O_3$ crystal by the vertical Bridgman method, Phys. Status Solidi B 262, 2400444 (2025). [DOI:10.1002/pssb.202400444]

[37] N. Xia, Y. Liu, D. Wu, L. Li, K. Ma, J. Wang, H. Zhang, and D. Yang, β-$Ga_2O_3$ bulk single crystals grown by a casting method, J. Alloys Compd. 935, 168036 (2023). [DOI:10.1016/j.jallcom.2022.168036]

[38] X. Gao, K. Ma, Z. Jin, D. Wu, J. Wang, R. Yang, N. Xia, H. Zhang, and D. Yang, Characteristics of 4-inch (100) oriented Mg-doped β-$Ga_2O_3$ bulk single crystals grown by a casting method, J. Alloys Compd. 987, 174162 (2024). [DOI:10.1016/j.jallcom.2024.174162]

[39] Y. Yan, D. Wu, N. Xia, T. Deng, H. Zhang, and D. Yang, Anisotropic thermal expansion tensor of β-$Ga_2O_3$ and its critical role in casting-grown crystal cracking, Appl. Phys. Lett. 124, 122102 (2024). [DOI:10.1063/5.0195733]

[40] A. Yoshikawa, V. Kochurikhin, T. Tomida, I. Takahashi, K. Kamada, Y. Shoji, and K. Kakimoto, Growth of bulk β-$Ga_2O_3$ crystals from melt without precious-metal crucible by pulling from a cold container, Sci. Rep. 14, 14881 (2024). [DOI:10.1038/s41598-024-65420-7]

[41] M. Kitahara, T. Tomida, V. Kochurikhin, G. Liudmila, K. Kamada, Y. Shoji, K. Kakimoto, and A. Yoshikawa, β-$Ga_2O_3$ crystal growth with cold container crucibles: Large-scale oxide crystal growth from cold crucible method, J. Cryst. Growth 677, 128461 (2026). [DOI:10.1016/j.jcrysgro.2025.128461]

[42] K. Kakimoto, I. Takahashi, T. Tomida, V. Kochurikhin, K. Kamada, S. Nakano, and A. Yoshikawa, Heat transfer in β-$Ga_2O_3$ crystal grown through a skull melting method, J. Cryst. Growth 629, 127553 (2024). [DOI:10.1016/j.jcrysgro.2023.127553]

[43] A. Lang and A. Makepeace, Synchrotron x-ray reticulography: principles and applications, J. Phys. D: Appl. Phys. 32, A97 (1999). [DOI:10.1088/0022-3727/32/10A/321]

[44] A. Lang and A. Makepeace, Reticulography: a simple and sensitive technique for mapping misorientations in single crystals, J. Synchrotron Rad. 3, 313 (1996). [DOI:10.1107/S0909049596010515]

[45] Y. Yao, K. Hirano, Y. Sugawara, and Y. Ishikawa, Domain boundaries in ScAlMgO4 single crystal observed by synchrotron radiation X-ray topography and reticulography, Semicond. Sci. Technol. 37, 115009 (2022). [DOI:10.1088/1361-6641/ac974b]

[46] Y. Yao, Y. Ishikawa, and Y. Sugawara, X-ray diffraction and Raman characterization of β-$Ga_2O_3$ single crystal grown by edge-defined film-fed growth method, J. Appl. Phys. 126, 205106 (2019). [DOI:10.1063/1.5129226]

[47] Y. Yao, Y. Ishikawa, and Y. Sugawara, Slip planes in monoclinic β-$Ga_2O_3$ revealed from its {010} face via synchrotron X-ray diffraction and X-ray topography, Jpn. J. Appl. Phys. 59, 125501 (2020). [DOI:10.35848/1347-4065/abc1aa]

[48] Y. Yao, Y. Ishikawa, Y. Sugawara, Y. Takahashi, and K. Hirano, Observation of threading dislocations in ammonothermal gallium nitride single crystal using synchrotron X-ray topography, J. Electron. Mater. 47, 5007 (2018). [DOI:10.1007/s11664-018-6252-3]

[49] Y. Yao, K. Hirano, Y. Takahashi, Y. Sugawara, K. Sasaki, A. Kuramata, and Y. Ishikawa, Visualization of the curving of crystal planes in β-$Ga_2O_3$ by X-ray topography, J. Cryst. Growth 576, 126376 (2021). [DOI:10.1016/j.jcrysgro.2021.126376]

[50] G. Green, N. Loxley, and B.K. Tanner, Dislocation images in X-ray section topographs of curved crystals, J. Appl. Crystallogr. 24, 304 (1991). [DOI:10.1107/s0021889891001619]

[51] V. Lider, X-ray diffraction topography methods (Review), Phys. Solid State 63, 189 (2021). [DOI:10.1134/s1063783421020141]

[52] E. Suvorov, X-ray topography: yesterday, today, and prospects for the future, J. Surf. Investig.: X-Ray, Synchrotron Neutron Tech. 12, 835 (2018). [DOI:10.1134/s1027451018050026]

[53] H. Yamaguchi, A. Kuramata, and T. Masui, Slip system analysis and x-ray topographic study on β-$Ga_2O_3$, Superlattice Microst. 99, 99 (2016). [DOI:10.1016/j.spmi.2016.04.030]

[54] G. Schwuttke, Direct observation of imperfections in semiconductor crystals by anomalous transmission of X rays, J. Appl. Phys. 33, 2760 (1962). [DOI:10.1063/1.1702544]

[55] K. Ishiji, T. Fujii, T. Araki, and T. Fukuda, Observation of defect structure in ScAlMgO4 crystal using X-ray topography, J. Cryst. Growth 580, 126477 (2022). [DOI:10.1016/j.jcrysgro.2021.126477]

[56] T. Fukuda, Y. Shiraishi, T. Nanto, T. Fujii, K. Sugiyama, R. Simura, H. Iechi, K. Tadatomo, and T. Matsuoka, Growth of bulk single crystal ScAlMgO4 boules and GaN films on ScAlMgO4 substrates for GaN-based optical devices, high-power and high-frequency transistors, J. Cryst. Growth 574, 126286 (2021). [DOI:10.1016/j.jcrysgro.2021.126286]

[57] M. Imai, Y. Shiraishi, M. Shibata, H. Noda, and Y. Yatsurugi, Quantitative measuring method of growth striations in Czochralski-grown silicon crystal, J. Electrochem. Soc. 135, 1779 (1988). [DOI:10.1149/1.2096129]

[58] Y. Yao, K. Hirano, H. Yamaguchi, Y. Sugawara, N. Okada, K. Tadatomo, and Y. Ishikawa, A synchrotron X-ray topography study of crystallographic defects in ScAlMgO4 single crystals, J. Alloys Compd. 896, 163025 (2021). [DOI:10.1016/j.jallcom.2021.163025]

[59] D. Hull and D.J. Bacon, Introduction to dislocations, 5th edition, Elsevier Ltd., Oxford, (2011) p.24–p.27

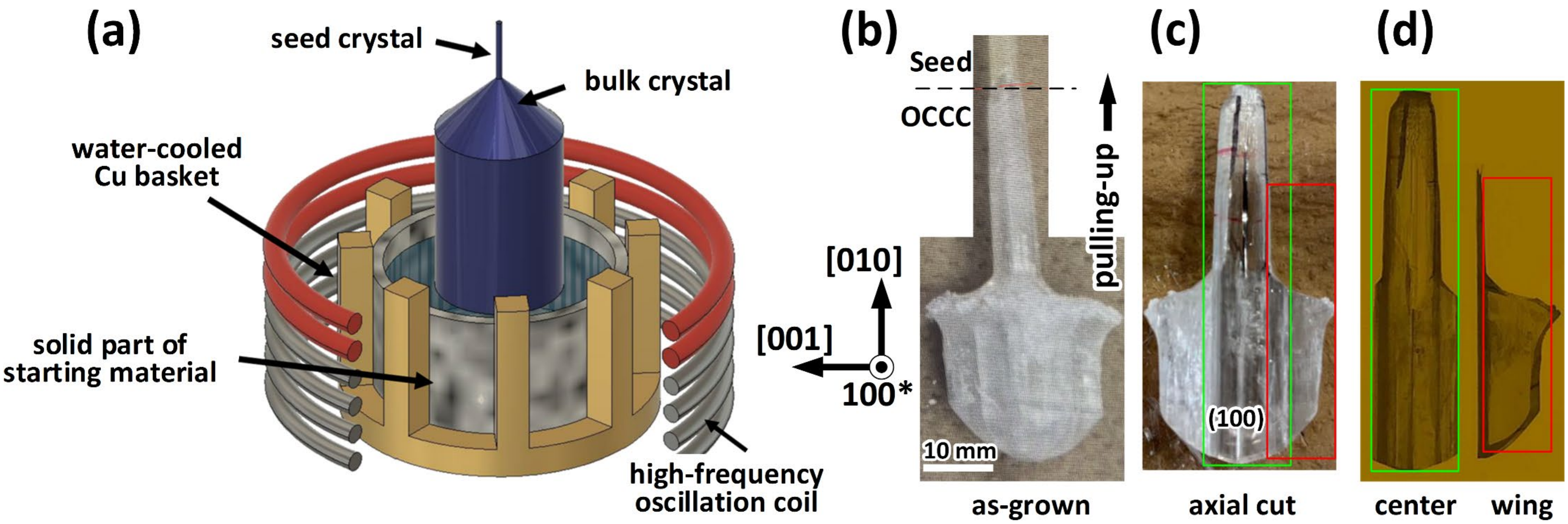

**Figure 1.** (a) Schematic illustration of the OCCC growth apparatus; (b) Optical photograph of the as-grown crystal; (c) View of the crystal from the (100) face after cleaving along the central axis; (d) Optical microscope images obtained in transmission mode.

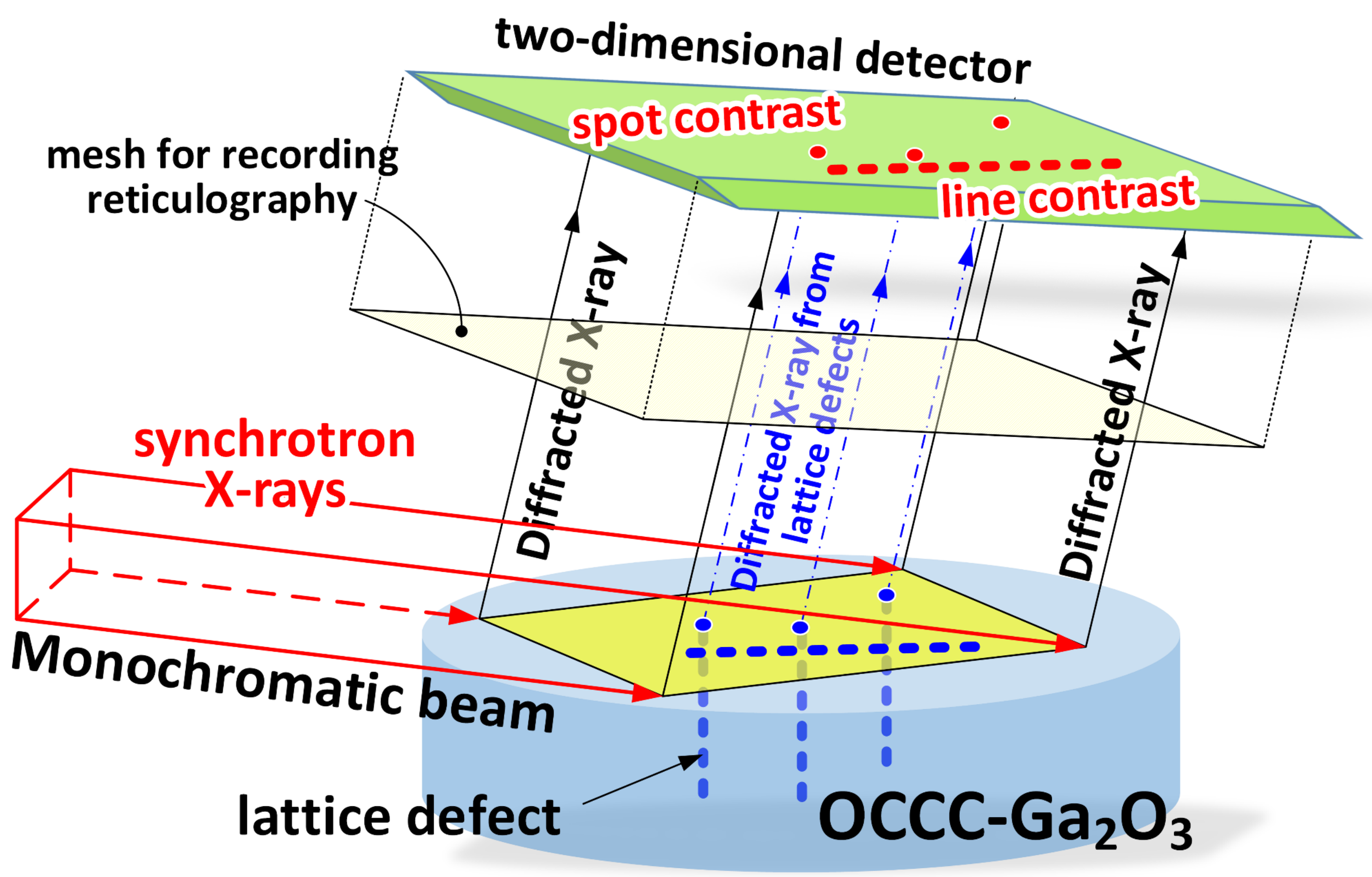

**Figure 2.** Schematic illustration of the experimental setup for X-ray topography and X-ray reticulography.

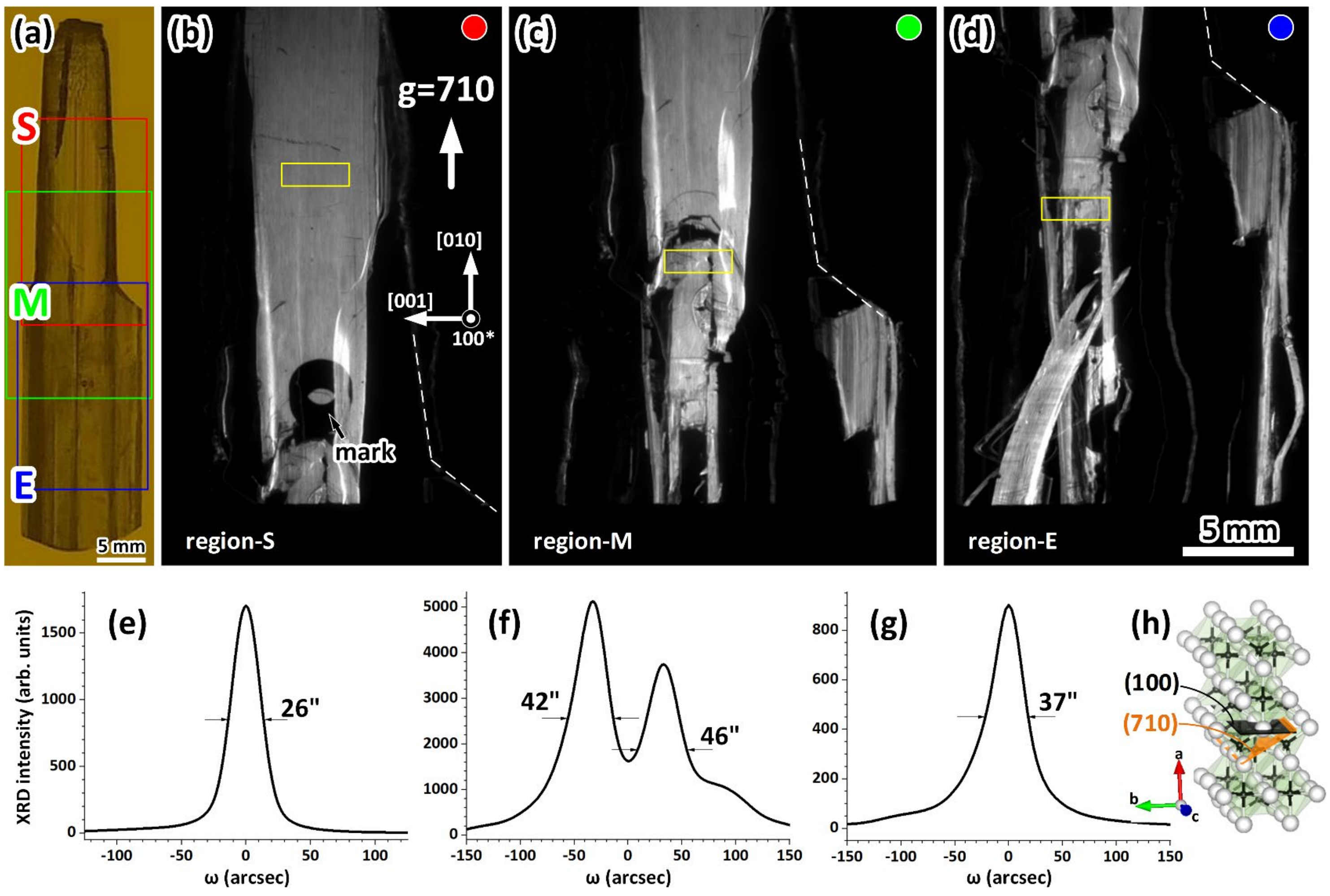

**Figure 3.** (a) Optical image of the central portion of the OCCC crystal containing regions S (start), M (middle), and E (end); (b)–(d) Maximum-intensity images constructed from sequences of XRDI data for regions S, M, and E, respectively, taken with $\boldsymbol{g}$ = 710; (e)–(g) XRD ω-rocking curves of regions S, M, and E, respectively. The integrated intensity of all pixels within each selected region was plotted as a function of ω to obtain the corresponding rocking curve; (h) Schematic illustration of the (710) lattice plane.

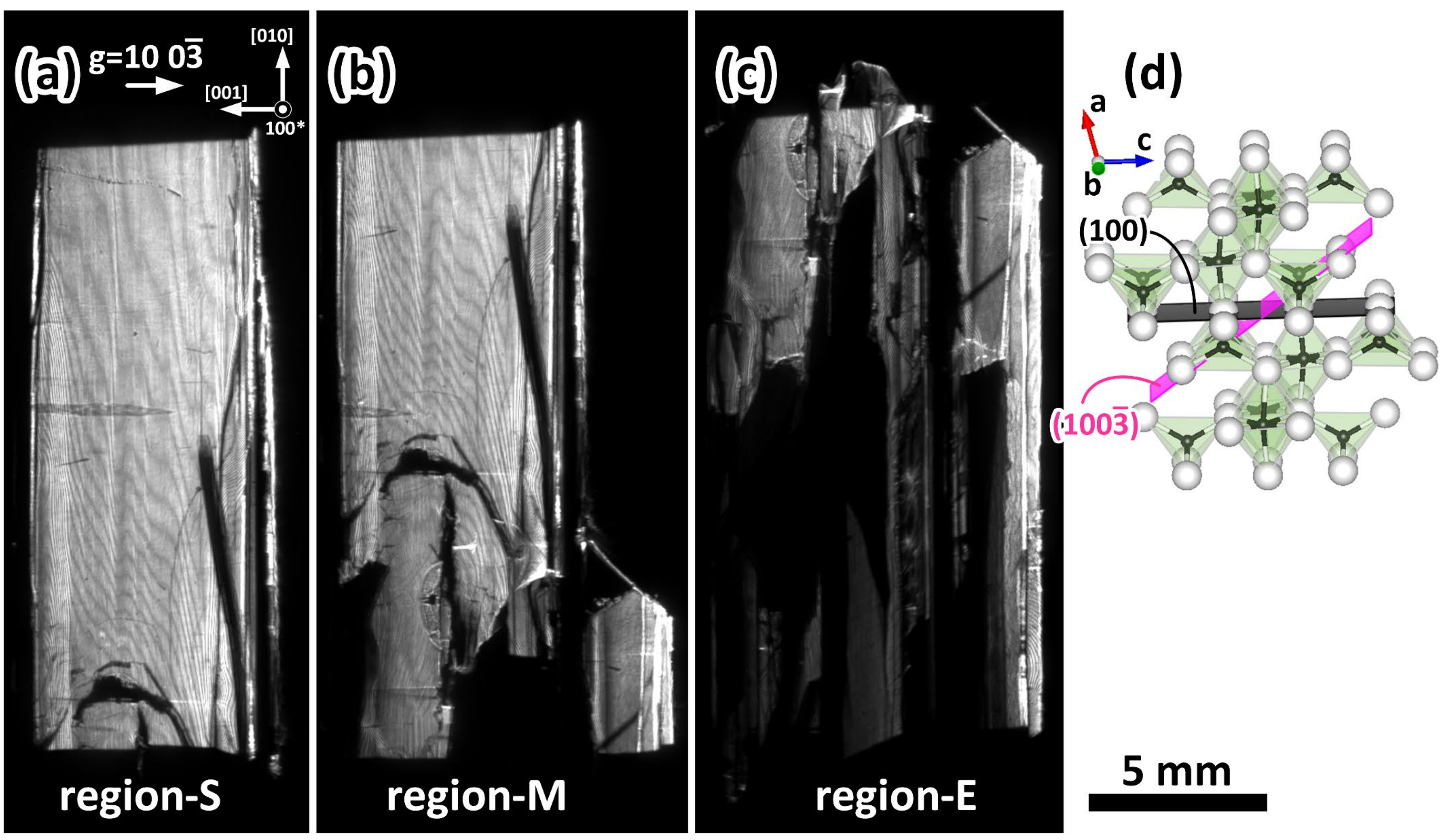

**Figure 4.** (a)–(c) Maximum-intensity images constructed from sequences of XRDI data for regions S, M, and E, respectively, taken with $\boldsymbol{g}$ = 10 0$\bar{3}$.; (d) Schematic illustration of the (10 0$\bar{3}$) lattice plane.

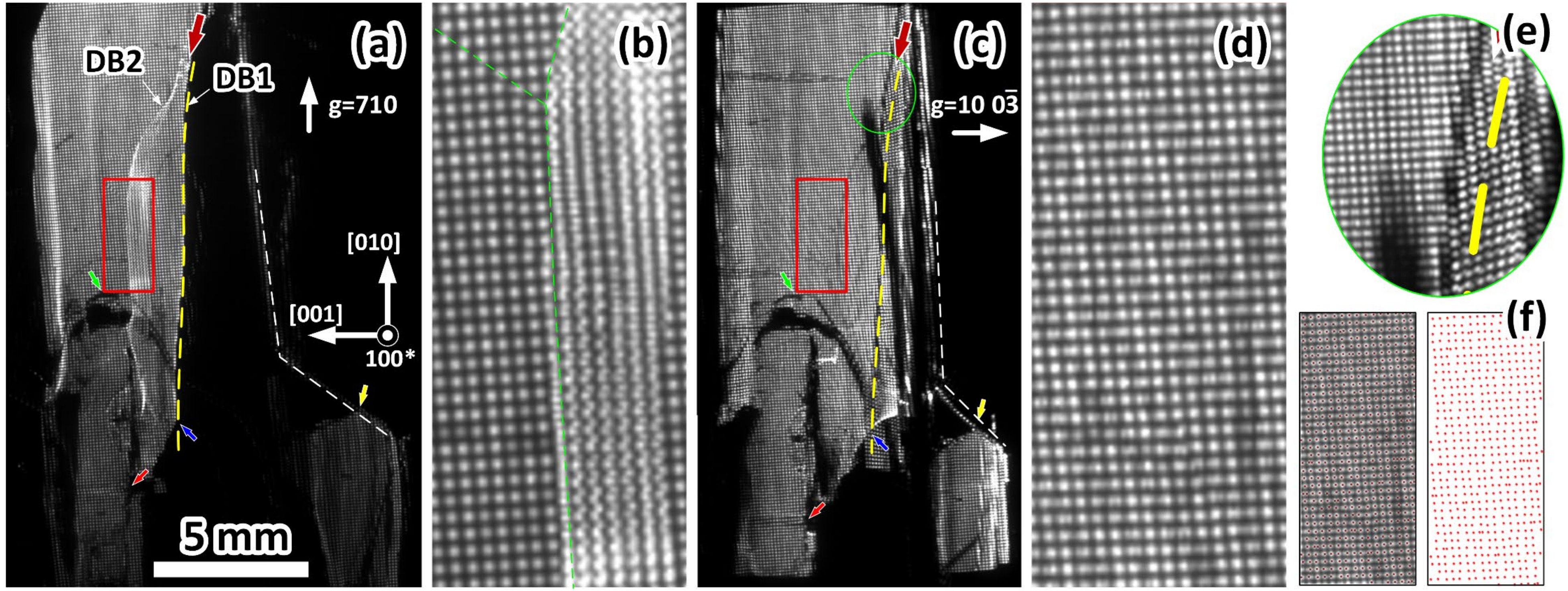

**Figure 5.** (a) X-ray reticulography image taken with $\boldsymbol{g}$ = 710 and (b) a magnified image of the area marked by the red frame; (c) X-ray reticulography image taken with $\boldsymbol{g}$ = 10 0$\bar{3}$ and (d) a magnified image of the area marked by the red frame; (e) Magnified image of the area marked by the green circle in Figure 5(c); (f) Coordinate analysis of the spot matrix in the red frame of the reticulography image in Figure 5(c).

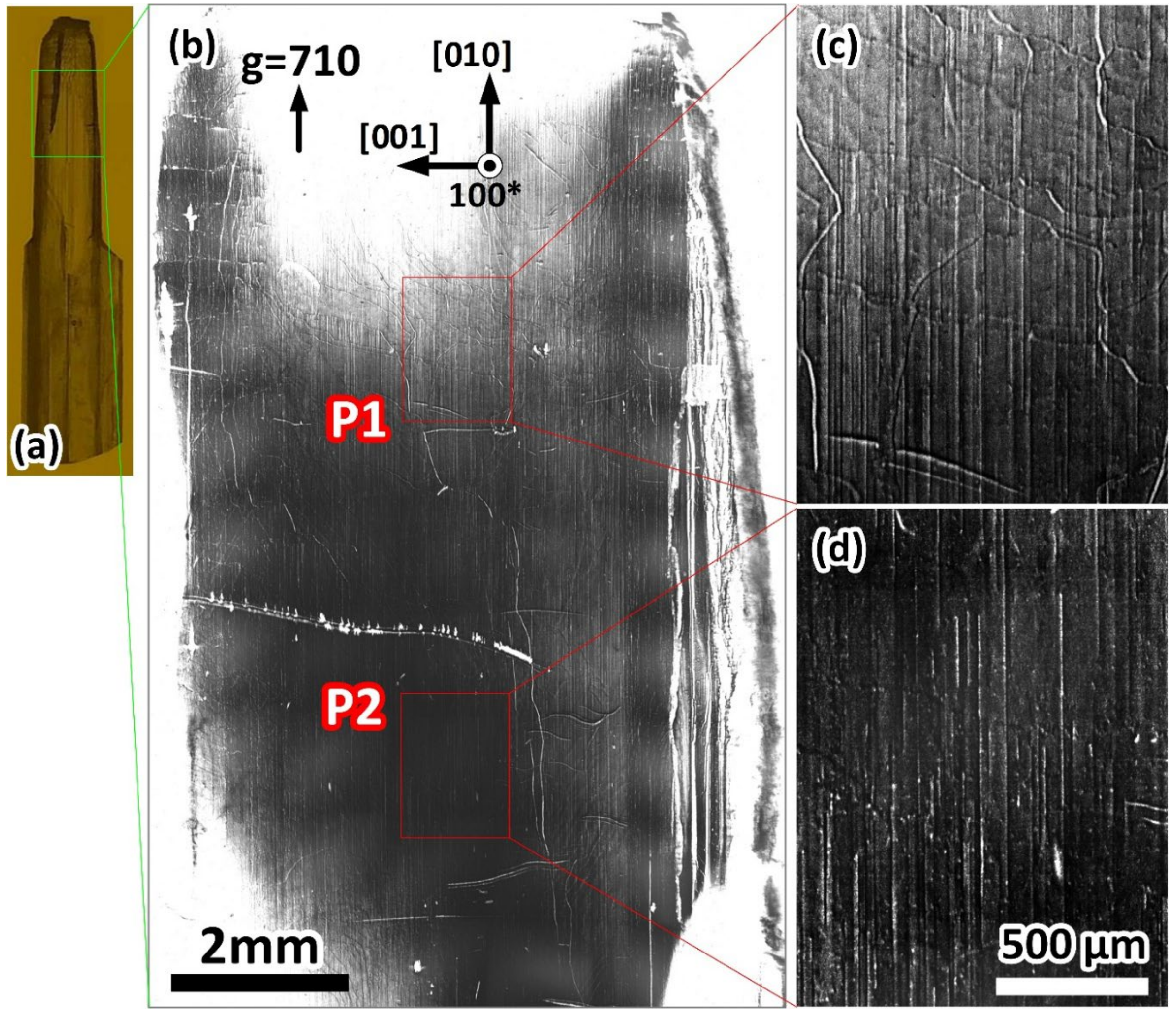

**Figure 6.** (a) Optical image of the central portion of the OCCC crystal. The area marked by the green frame belongs to region S; (b) XRT image recorded on X-ray film using $\boldsymbol{g}$ = 710; (c), (d) Magnified images corresponding to regions P1 and P2, respectively. The XRT image was recorded on an X-ray film using a single exposure under fixed diffraction conditions.

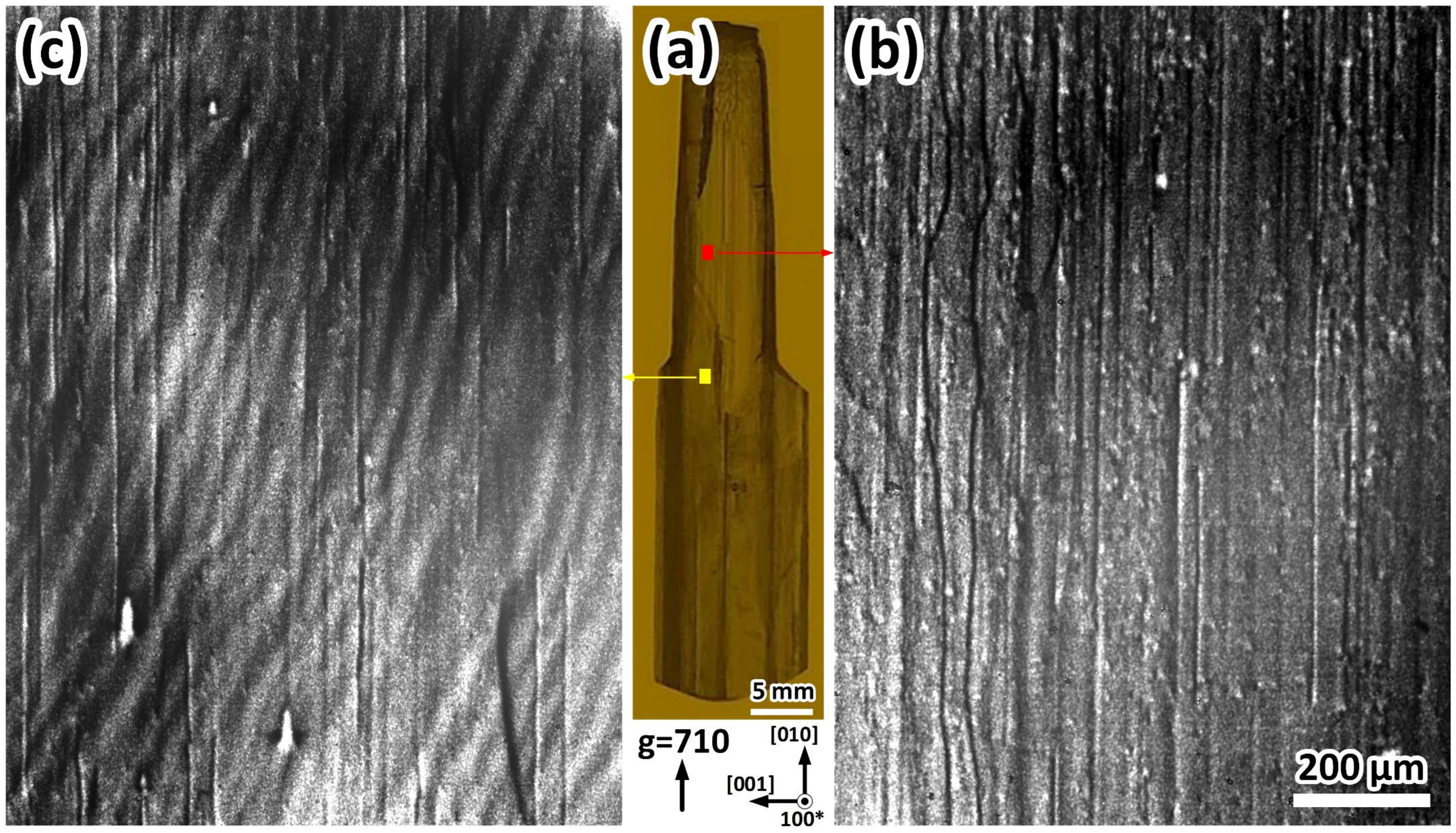

**Figure 7.** (a) Optical image of the central portion of the OCCC crystal. The areas indicated by the red and yellow marks belong to region M; (b), (c) XRT images recorded on X-ray film using $\boldsymbol{g}$ = 710.

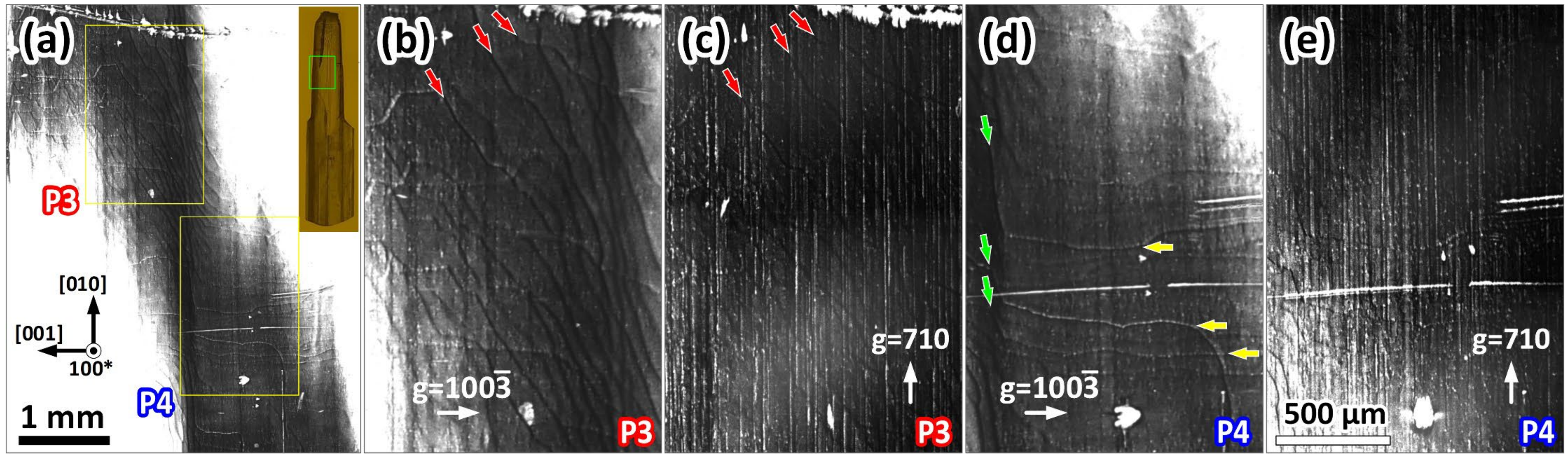

**Figure 8.** (a) XRT image selected from an area in region S; (b), (c) XRT images of region P3 taken with $\boldsymbol{g}$ = 10 0$\bar{3}$ and $\boldsymbol{g}$ = 710, respectively; (d), (e) XRT images of region P4 taken with $\boldsymbol{g}$ = 10 0$\bar{3}$ and $\boldsymbol{g}$ = 710, respectively.

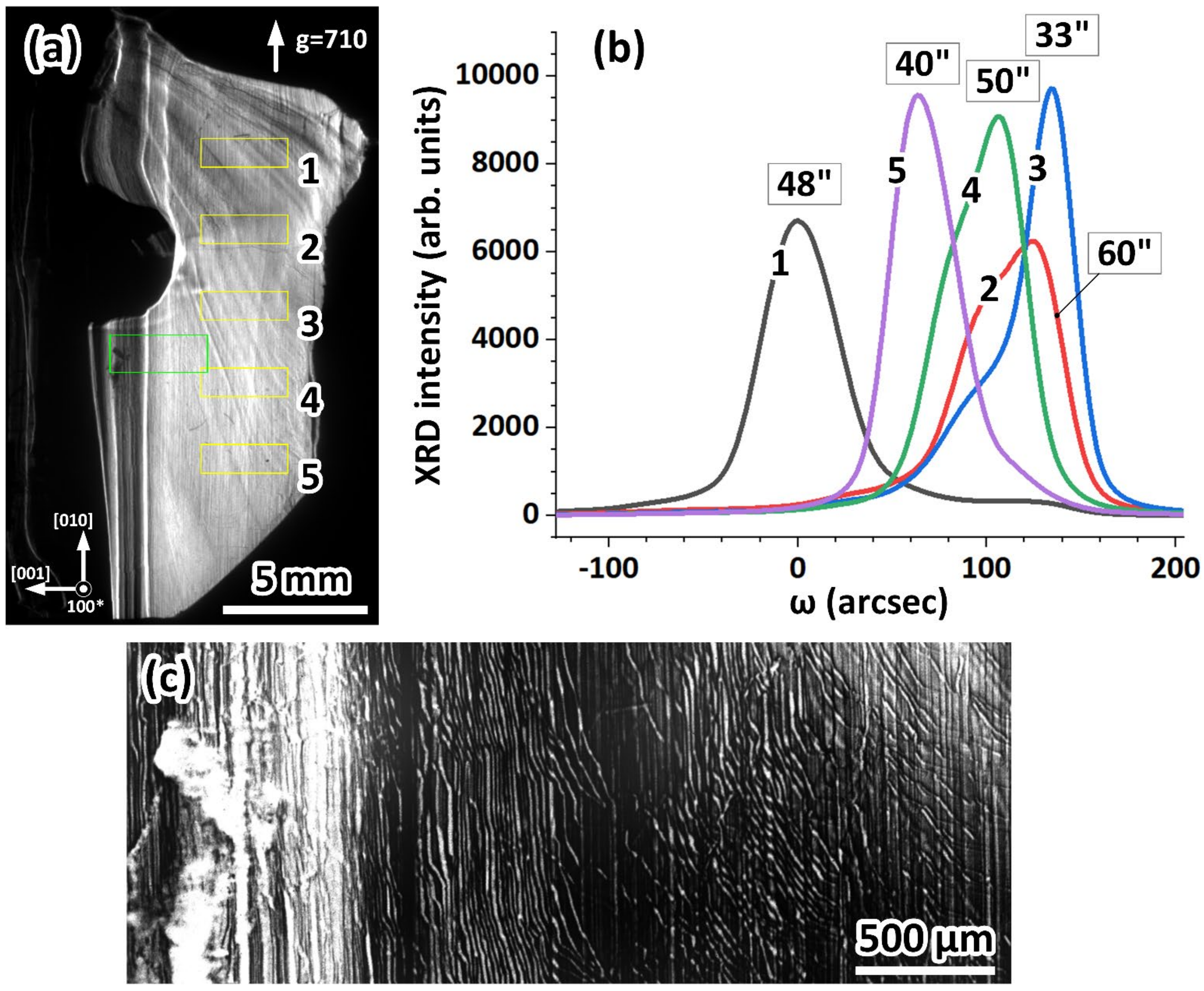

**Figure 9.** (a) Maximum-intensity image constructed from a sequence of XRDI data for the wing portion, taken with ***g*** = 710; (b) XRD ω-rocking curves for regions 1–5, whose locations are indicated in Figure 9(a); (c) XRT image of the region indicated by the green frame in Figure 9(a).